\documentclass{emulateapj}
\usepackage{graphicx}
\usepackage{natbib}

\singlespace

\shortauthors{Kogut and Fixsen}
\shorttitle{Foreground Bias From Parametric Models of Far-IR Dust Emission}
\slugcomment{Accepted for publication in The Astrophysical Journal}

\begin{document}

\title{Foreground Bias From Parametric Models of Far-IR Dust Emission}

\author{ 
A. Kogut\altaffilmark{1},
D.\ J.\ Fixsen\altaffilmark{2,1}
}

\altaffiltext{1}{Code 665, Goddard Space Flight Center, Greenbelt, MD 20771}
\email{Alan.J.Kogut@nasa.gov}
\altaffiltext{2}{University of Maryland, College Park, MD, 20742}


\begin{abstract}
We use simple toy models of far-IR dust emission
to estimate the accuracy to which
the polarization of the cosmic microwave background
can be recovered using multi-frequency fits,
if the parametric form chosen for the fitted dust model
differs from the actual dust emission.
Commonly used approximations to the far-IR dust spectrum
yield CMB residuals comparable to or larger than the sensitivities
expected for the next generation of CMB missions,
despite fitting the combined CMB $+$ foreground emission
to precision 0.1\% or better.
The Rayleigh-Jeans approximation to the dust spectrum
biases the fitted dust spectral index by $\Delta \beta_d = 0.2$
and the inflationary B-mode amplitude by $\Delta r = 0.03$.
Fitting the dust to a modified blackbody at a single temperature
biases the best-fit CMB by $\Delta r > 0.003$
if the true dust spectrum 
contains multiple temperature components.
A 13-parameter model fitting two temperature components
reduces this bias by an order of magnitude
if the true dust spectrum is in fact a simple superposition
of emission at different temperatures,
but fails at the level $\Delta r = 0.006$
for dust whose spectral index varies with frequency.
Restricting the observing frequencies to a narrow region
near the foreground minimum reduces these biases for some dust spectra
but can increase the bias for others.
Data at THz frequencies surrounding the peak of the dust emission
can mitigate these biases while providing a direct determination
of the dust temperature profile.
\end{abstract}



\section{Introduction}
Polarization of the cosmic microwave background (CMB)
provides a critical test for models of inflation.
The primary signature is is a parity-odd curl component
in the polarization on angular scales of a few degrees
or larger
\citep{kamionkowski/etal:1997,
seljak/zaldarriaga:1997}.
The amplitude of this ``B-mode'' signal
depends on the inflationary potential
\begin{equation}
V^{1/4} = 1.06 \times 10^{16} ~{\rm GeV} 
\left( 
\frac{r}{0.01} 
\right)^{1/4}
\label{potential_eq}
\end{equation}
where 
$r$ is the power ratio of the tensor (gravitational) perturbations
to scalar (density) fluctuations
\citep{turner/white:1996}.
If inflation results from Grand Unified Theory physics
(energy $\sim 10^{16}$ GeV),
the B-mode amplitude should be in the range 1 to 100 nK.
Signals at this amplitude
could be detected by  a dedicated polarimeter,
providing a critical test of a central component of modern cosmology.

\begin{figure}[t]
\vspace{-2mm}		
\includegraphics[width=2.6in, angle=90]{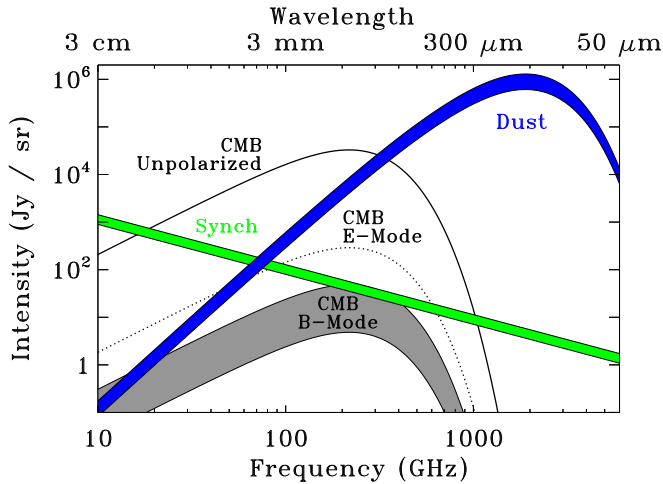}
\caption{Frequency spectra of the CMB and polarized foregrounds.
The grey band shows 0.01$<$$r$$<$0.1  
for the primordial inflation signal.
Colored bands show the synchrotron and dust foregrounds
for the cleanest 50\% and 75\% of the sky.
The inflationary signal is fainter than Galactic foregrounds
at all frequencies, 
requiring accurate models for foreground subtraction.
}
\label{foreground_fig}
\end{figure}

Detecting the inflationary signal will be challenging.
A primary concern is
confusion from astrophysical foregrounds.
We view the CMB through a screen of diffuse Galactic emission
originating within different components of the interstellar medium.
Figure \ref{foreground_fig} 
compares the inflationary B-mode signal to polarized Galactic foregrounds.
Synchrotron emission from relativistic cosmic ray electrons
accelerated in the Galactic magnetic field
dominates the diffuse radio continuum
at low frequencies.
For a power-law distribution of cosmic ray energy
$N(E) \sim E^{-p}$,
the synchrotron intensity is also a power law,
\begin{equation}
I_s(\nu) = A_s \left(\frac{\nu}{\nu_s} \right)^{\beta_s},
\label{synch_power_law}
\end{equation}
\noindent
where 
$\nu$ is the observing frequency,
$\beta_s = (1-p)/2$
is the spectral index, and
$A_s$
is the amplitude
defined relative to reference frequency~$\nu_s$
\citep{rybicki/lightman:1979}.
The measured values 
$2.6 < p < 3.2$
for the cosmic ray energy spectrum
correspond to synchrotron spectral index
$-1.1 < \beta_s < -0.8$,
in reasonable agreement with radio data
\citep{strong/etal:2007,
jaffe/etal:2011,
kogut:2012,
bennett/etal:2013,
planck:2015:10}.
The synchrotron spectrum may contain additional features
(curvature, spectral break),
but evidence for such features is restricted to frequencies below 20 GHz
and is not considered here.

Dust is the dominant foreground at high frequencies.
Dust grains in the interstellar medium
absorb optical and UV photons
and re-radiate the energy in the far-infrared.
The resulting spectrum is often
empirically modeled 
as a sum of modified blackbodies with power-law emissivities,
\begin{equation}
I_d(\nu) = \sum_i \epsilon_i B_\nu(T_i) 
\left( \frac{\nu}{\nu_d} \right)^{\beta_{d\,i}}
\label{dust_greybody_eq}
\end{equation}
where
$B_\nu(T)$ is the Planck intensity at temperature $T$,
and
the emissivity $\epsilon$
and spectral index $\beta_d$
are defined at reference frequency $\nu_d$.
Fitting the high-latitude dust cirrus to a single modified blackbody
returns values
$T_d~=~20$ K
and
$\beta_d = 1.6$
\citep{planck:2015:22}.	

The B-mode signal is fainter than the Galactic foregrounds
at all frequencies.
Current measurements limit primordial B-modes to amplitude 
$r < 0.07$ at 95\% confidence
\citep{bicep/planck:2015}.
Distinguishing primordial B-modes from Galactic foregrounds
at this level
requires subtracting the foregrounds to few percent accuracy or better.
The next generation of CMB polarimeters
anticipates sensitivities $r < 0.001$.
Measurements at this level
require foreground subtraction 
with sub-percent accuracy.

Despite their importance,
diffuse Galactic foregrounds at millimeter wavelengths
are poorly constrained.
The observed dust emission depends on a number of factors
including the 
grain size distribution,
chemical and physical composition of the dust grains,
competing emission mechanisms within a grain,
and the three-dimensional distribution
of the dust population
irradiated by the stellar UV/optical field.
None of these are known in detail.
Lacking a detailed physical model,
CMB analyses typically use a 
purely phenomenological model for the dust,
treating it as the sum of one or two modified blackbody components
along each line of sight.
In this paper, 
we use simple extensions 
to commonly used models of far-IR dust emission
to estimate the systematic error (bias)
in the B-mode amplitude $r$
due to the use of such phenomenological models.

\section{Dust Emission Models}
A common approximation
treats far-IR dust emission as a modified blackbody
at a single well-defined temperature $T_d$
(Eq. \ref{dust_greybody_eq}).
Applied to the Planck polarization data,
the single-temperature model
yields dust temperature $19.6 \pm 0.8$ K 
with spectral index $1.59 \pm 0.02$
\citep{planck:2015:22}.
However, data at higher frequencies
show that roughly 4\% of the far-IR power
resides in a component
consistent with a colder temperature near 10 K
\citep{wright/etal:1991,
reach/etal:1995,
dwek/etal:1997,
finkbeiner/etal:1999,
meisner/finkbeiner:2015}.
The origin of this component is unknown.
The sum of two blackbodies
can not be modeled to arbitrary precision 
as a third blackbody:
the single-temperature model is at best an approximation
to the underlying dust spectrum.

The two-component modified blackbody model fits the far-IR dust intensity
along each line of sight
to the superposition of emission 
from components at two discrete temperatures.
This, too, can only approximate the underlying spectrum.
The dust emission spectrum represents a balance
between optical/UV power absorbed by each grain
and subsequently re-radiated in the far-infrared.
It thus depends on both the absorption and emission properties
of individual dust grains
as well as the grain size and
the stellar radiation field illuminating each grain.
All of these will vary within the interstellar medium.
The stellar radiation field {\it must} vary 
as individual dust grains lie at different distances 
from individual stellar sources.
Differences in the physical properties of dust grains
yield additional temperature differences,
with carbonaceous grains systematically colder than silicates
(see, e.g.,
\citet{zubko/etal:2004,
draine/li:2007,
draine/fraisse:2009}).
Transient heating of small grains
drives further temperature variation,
even within a homogeneous dust population
illuminated by a uniform stellar radiation field.

\begin{figure}[b]
\centerline{ 
\includegraphics[width=3.2in]{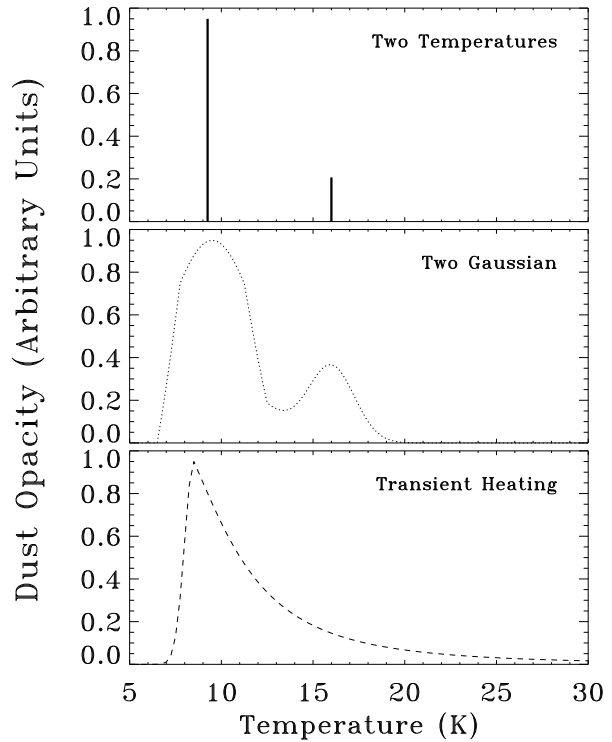}
}
\caption{Far-IR dust opacity vs temperature
for several toy models of far-IR dust emission.
(Top) Two-component modified blackbody emission
for which the dust opacity is non-zero
at exactly two temperatures.
(Middle) A broader distribution
with two peaks smoothed by a Gaussian distribution.
(Bottom) Distribution with a single low-temperature peak
and an extended ``tail'' toward higher temperatures,
approximating a population with transient heating.
}
\label{tau_vs_temp}
\end{figure}

Systematic differences
between the spectrum of the diffuse dust cirrus
and the parametric models
used to describe this emission
will bias foreground subtraction for CMB observations.
Models of foreground emission 
sufficient for E-mode or unpolarized analyses
can prove inadequate at the higher accuracy
required for B-mode analysis.
We quantify the 
extent to which specific choices for the dust parameterization
bias the fitted CMB terms
using toy models of dust emission.
We begin with simple models 
of far-IR emission
that have been used for CMB fitting
(Rayleigh-Jeans approximation,
one- and two-component modified blackbodies),
then 
extend the  parameterization
to include a distribution in dust temperature
(Figure \ref{tau_vs_temp}).
The simplest extension beyond the two-temperature model
retains two temperature peaks,
but broadens the distribution
using a Gaussian smoothing
to include emission at intermediate temperatures as well.
We additionally consider a model
with a single peak at low temperatures
plus an extended ``tail'' to higher temperatures
to approximate the effect of transient heating.
For both the Gaussian and transient models,
we adjust the relative amplitudes of the warm and cold components
so that emission from the cold component
remains at the observed 3.7\% of the power
from the warm component
while forcing the combined dust emission
to match the intensity of Planck polarized dust model at 353 GHz.
Finally, we consider a model in which the
dust spectral index varies with frequency,
flattening the spectrum at lower frequencies.
Such ``running'' of the spectral index
is an expected characteristic of
two-level emission in disordered dust grains
\citep{meny/etal:2007,
paradis/etal:2011}.

Although not intended to represent in detail
any specific physical model of far-IR dust emission,
the toy models used here
capture sufficient features 
of leading models
to examine the implications for
simple parametric dust models.
Other forms for the dust emission are possible.
So-called ``anomalous microwave emission''
is a significant contribution to the
unpolarized intensity at frequencies below 90 GHz
\citep{kogut/etal:1996,
gold/etal:2011,
planck.2011.20,
genova-santos/etal:2015}
and is thought to result from a population
of small, spinning dust grains
\citep{draine/lazarian:1998,
ali-hamoud/etal:2009,
hoang/etal:2010}.
Anomalous microwave emission
is not known to be polarized,
with upper limits 
to the fractional polarization 
at the sub-percent level
\citep{genova-santos/etal:2015}.
Dust containing magnetic material
could also emit through thermal fluctuations of the grain magnetization,
which would have comparable polarization
to thermal dust emission but
would differ significantly from a power-law emissivity
\citep{draine/hensley.2012}.

\section{Simulations}
Several authors have estimated the 
sensitivity and possible systematic bias in $r$
resulting from foreground contamination of CMB observations
\citep{fantaye/etal:2011,
remazeilles/etal.2016}.
These analyses 
only examine simple single-temperature dust models
and are limited
to choices of observing frequencies and instrumental sensitivity
matched to specific CMB missions.
Here we focus on the bias
resulting from the unknown dust spectra,
with a broader range of plausible dust emission models
not limited to instrument-specific observing frequencies.

We evaluate the CMB and foreground emission
using noiseless simulations.
All simulations adopt a power-law model (Eq. \ref{synch_power_law})
with $\beta_s = -1.05$
for the synchrotron spectral dependence\footnote{
Recall that the synchrotron spectral index
$\beta_s$ is specified
in units of intensity, not antenna temperature.}
%
but use different toy models for the dust.
Rather than simulate a specific instrument configuration,
we evaluate the simulations
using frequency channels spaced every 10 GHz
starting at 30 GHz
and extending to some maximum frequency 
(typically 500 GHz).
The number of frequency channels is thus much larger
than the number of fitted parameters
so that the results do not depend on details of the frequency basis.
We normalize the foreground emission
in Stokes $Q$ and $U$
to match the Planck polarized foreground maps
at 30 GHz for synchrotron emission
and 353 GHz for dust
\citep{planck:2015:10}.
The dust normalization is thus fixed by observation
and does not depend on assumptions for the fractional polarization
of the dust emission.
We degrade the maps
to a common {\tt HEALPIX} resolution {\tt NSIDE}=256
\citep{gorski/etal:2005}
and fit the superposed emission
in each pixel outside the Planck HFI polarization mask
to combination of CMB, synchrotron, and dust emission
where the parametric form of the fitted dust model
now differs from the input.
The fit minimizes the weighted sky residual
\begin{equation}
\Gamma^2 =   \sum_i
\left(
\frac
{ I_{\rm fit} - I_{\rm input} }
{W}
\right)^2_i .
\label{resid_eq}
\end{equation}
Although the input model has no noise contribution,
we weight each frequency channel
by the inverse of the combined input sky intensity,
$ W = I_{\rm input}$,
so that the fit is not dominated 
by channels where the foregrounds are much brighter than the CMB.
This approximates common practice for CMB observations,
where the sensitivity in each channel
is chosen
to produce equal signal-to-noise ratio
at all frequencies.
Other choices for channel weights are possible
(for instance, uniform weight at all frequencies).
The results do not depend sensitively upon the choice of weights.
The fitted $Q$ and $U$ amplitudes
for the CMB component in each pixel
are then used 
in a spherical harmonic analysis
to determine
the amplitude of the B-mode signal,
which we compare to the simulation input
to determine the bias in $r$.
Note that this automatically includes
the typical 2:1 ratio of E-mode to B-mode power
observed for the polarized dust foreground.

\begin{figure}[t]
\includegraphics[width=2.6in, angle=90]{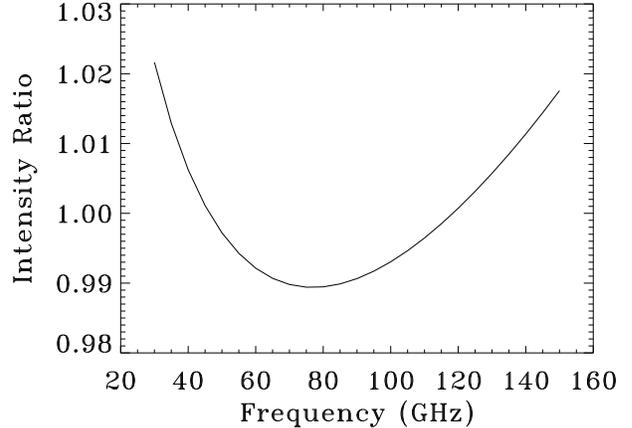}
\caption{Intensity ratio of power-law dust emission
normalized to a single-temperature modified blackbody
with $T_d = 20$K (see text).
The Rayleigh-Jeans approximation
fails at the few-percent level
for frequencies above 20 GHz
and dust temperatures 
appropriate for the diffuse dust cirrus. 
}
\label{rj_curvature}
\end{figure}

\vspace{15mm}

\section{CMB Bias from Parametric Dust Models}
A commonly used technique for foreground identification
and subtraction is parametric modeling,
in which measurements of the sky intensity 
in multiple frequency channels
are modeled as the sum of CMB and foreground components
specified by their frequency dependence.
We evaluate noiseless simulations
using different parametric forms
for the fitted dust model
to quantify the dependence of CMB residuals
on the choice of input versus fitted dust models.

\subsection{Rayleigh-Jeans Approximation}
At sufficiently low frequencies,
dust emission may be approximated 
as a power law in intensity,
\begin{equation}
I_d(\nu) = a_d \left( \frac{\nu}{\nu_d} \right)^{2 + \beta_{d}} ~.
\label{rj_approx}
\end{equation}
This Rayleigh-Jeans approximation 
has seen considerable use 
for modeling the dust contribution to CMB measurements
at frequencies from 22 GHz to 410 GHz
\citep{kogut/etal:1996,
masi/etal:2001,
ponthieu/etal:2005,
paladini/etal:2007,
dupac:2009,
bennett/etal:2013}.
Although appropriate for the sensitivity levels of these analyses,
reliance on the Rayleigh-Jeans approximation
will in general bias both the fitted dust parameters
and the estimated CMB amplitudes.
The Rayleigh-Jeans approximation is valid
in the limit 
$x~=~h \nu / k T_d \ll~1$
where
$h$ is Planck's constant,
$k$ is Boltzmann's constant,
and $T_d$ is the dust temperature.
For dust at temperature $T_d = 20$ K,
$x=0.24$ at 100 GHz
and does not fall below 0.05 until $\nu < 21$ GHz.
At the frequencies $\nu > 20$ GHz
used by most CMB missions,
the spectral curvature of the Planck function $B_\nu(T)$
can not be approximated by a power-law
to percent-level accuracy.
Figure \ref{rj_curvature} compares dust emission
from a single modified blackbody (Eq. \ref{dust_greybody_eq})
with $T_d = 20$ K and $\beta_d = 1.6$
to the best-fit Rayleigh-Jeans approximation.
The curvature of the modified blackbody 
relative to a power-law fit 
is apparent at the few-percent level.

\begin{figure}[t]
\includegraphics[width=3.2in]{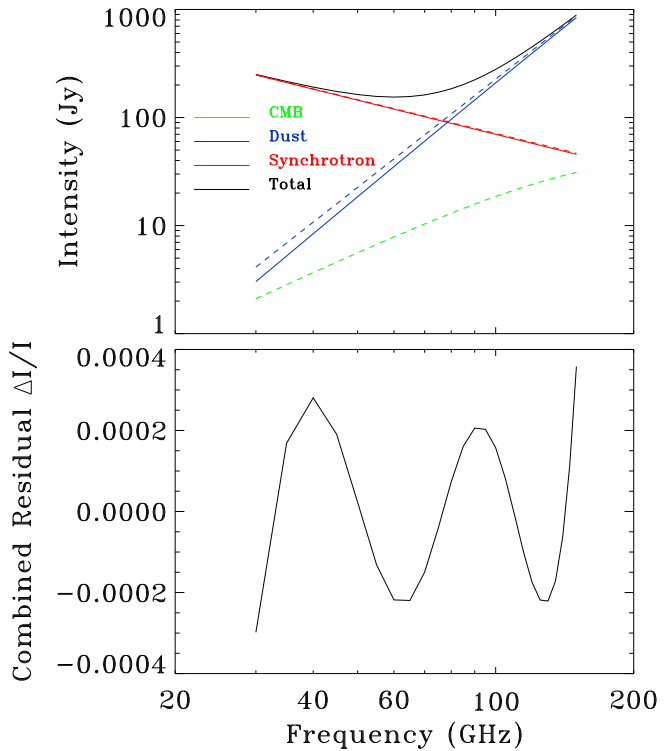}
\vspace{-5mm}
\caption{Input spectra
compared to best-fit model
when the input dust consists of a single modified blackbody
but the fitted model assumes 
the power-law Rayleigh-Jeans approximation.
(Top) Input spectra (solid lines)
and best-fit model (dashed lines).
The input has synchrotron and dust but no CMB,
with foreground amplitudes
at the median brightness of the polarized Planck foreground model.
(Bottom) Fractional residual
$\Delta I / I$ for the combined spectra.
Although the multi-component fit
matches the combined sky spectrum
to sub-percent precision,
the Rayleigh-Jeans approximation
biases the recovered amplitudes and spectral index values,
generating a false CMB term at level $r \sim 0.03$.
}
\label{rj_fit}
\end{figure}

Use of the Rayleigh-Jeans approximation
in multi-component foreground estimation
will bias estimates of the individual components.
Figure \ref{rj_fit} compares
the input and fitted sky models
for noiseless simulations ($\S$3)
where the input dust 
consists of a modified blackbody with $T_d = 19.6$ K and $\beta_d = 1.59$.
Fitting the simulated sky 
over the frequency range 30--140 GHz
using the Rayleigh-Jeans approximation
for the dust component
matches the superposed input spectra
to precision $\Delta I / I < 0.04$\%,
but biases both the fitted dust and CMB components.
The differential curvature of the
modified blackbody dust input
compared to the power-law output model
over-estimates the dust emission by 15\% 
at frequencies near the foreground minimum at 70 GHz.
The best-fit spectral indices are displaced from the input values,
with fitted values
$\beta_s = -1.04$ for synchrotron
and
$\beta_d = 1.33$ for dust.
The shift in dust spectral index
produced by the Rayleigh-Jeans approximation at low frequencies
mimics the flattening seen in the two-level-system model
although in this case it is entirely an artifact of the approximation.
The fitted model returns a false CMB term
with (negative) amplitude 80 nK,
comparable to a B-mode signal at level $r = 0.03$
for the multipole range 
$50 < \ell < 200$ 
targeting the ``recombination bump.''

\subsection{One-component Modified Blackbody}
The simplest model beyond the Rayleigh-Jeans approximation
treats the dust spectrum as modified blackbody
at a single temperature
with power-law emissivity.
A single-temperature model
allows fits across a wide frequency range
but can yield biased results
if the far-IR dust emission differs systematically
from the model.

\begin{figure}[t]
\includegraphics[width=2.7in,angle=90]{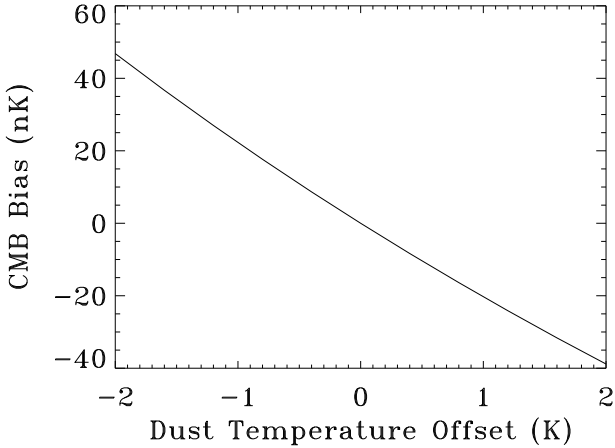}
\caption{Bias in fitted CMB component
when both the input dust and the fitted model
follow a single-temperature modified blackbody parameterization,
but the fitted dust temperature
differs from the true dust temperature.
}
\label{grebody_vs_dt}
\end{figure}

One such effect,
present even if the dust were dominated by just one temperature component,
results from errors in the dust temperature.
Measurements at millimeter wavelengths 
may lack either the frequency coverage
or sensitivity
for direct estimation of the dust temperature in each pixel,
relying instead on measurements
at higher frequencies near the peak dust intensity
or on spatial averaging across larger regions on the sky.
Data from the FIRAS, DIRBE, and {\it Planck} missions
show spatial variation of 3.5 K in the dust temperature across the sky
\citep{finkbeiner/etal:1999,
planck:2015:10}.
Substituting a mean dust temperature
into multi-frequency foreground models
will bias the fitted components
if the dust temperature within any pixel
differs from the supplied mean value.
Figure \ref{grebody_vs_dt} illustrates the effect.
As before,
we fit a noiseless superposition 
of synchrotron and dust emission with no CMB term,
evaluated at frequencies 30--500 GHz
using a model consisting of synchrotron, dust, and CMB.
The input dust model consists of
a single modified blackbody
with $T_d = 19.6$K and $\beta_d = 1.59$.
The fitted dust model
uses the same modifed blackbody parameterization
as the input,
but with dust temperature $T_d$ 
offset by $\Delta T$
with respect to the input model.
The fitted parameters are thus
the synchrotron amplitude and spectral index,
the dust amplitude and spectral index,
and the CMB amplitude.
Differences between the true dust temperature
and the temperature assumed for the model
result in a biased CMB component,
with a 1K dust temperature offset
corresponding to a CMB amplitude of 20 nK.
Parametric fits that use an externally supplied value 
for the dust temperature
generate a bias in the
fitted CMB amplitude
if the assumed dust temperature in each pixel
differs from the true value.
Keeping this term
below the 3 nK amplitude corresponding to $r = 0.001$
requires knowledge of the dust temperature
in each pixel accurate to 0.15 K or better.

\subsection{Two-Component Modified Blackbody}
Data at THz frequencies where dust emission peaks
show that 
3.7\% of the far-IR dust power 
is contained in a second, colder component
at temperature 8--12 K
\citep{reach/etal:1995,
finkbeiner/etal:1999,
meisner/finkbeiner:2015}.
The superposition of two blackbodies at different temperatures
cannot be modeled to arbitrary accuracy
as a single blackbody at a third temperature.
The existence of a cold component 
in the interstellar dust cirrus
will necessarily bias CMB fits
that ignore this component.

\begin{figure}[t]
\includegraphics[width=2.7in,angle=90]{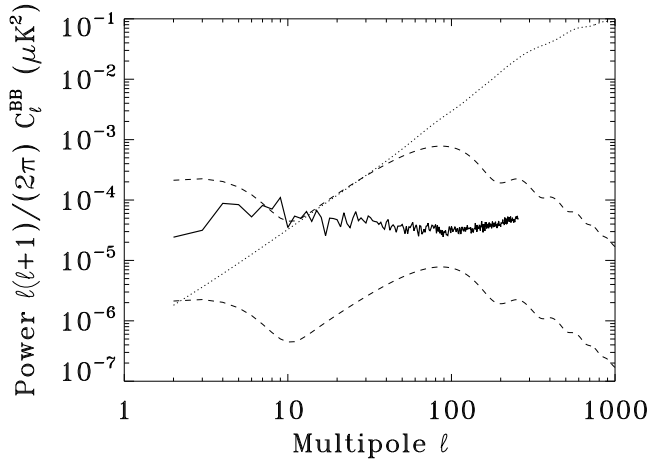}
\caption{Power spectrum of CMB residuals
created when a sky with modified blackbody dust
with two temperature components
is fit using a single-temperature modifed blackbody model.
The thin dashed lines show inflationary models
with $r=0.01$ and $r=0.001$
while the thin dotted line shows the lensing signal.
Fitting two-tempeature dust with a single-temperature model
creates CMB residuals large compared to $r=0.001$.
}
\label{2_greybody_cl}
\end{figure}

We evaluate the effect by fitting a simulated sky spectrum
over the frequency range 30 -- 500 GHz
using a superposition of synchrotron and dust emission
with no CMB term.
The dust emission in each pixel
is a superposition of two modified blackbody components:
a warm component with temperature $T_d \sim 16$ K 
and spectral index $\beta_d = 2.70$,
plus a cold component with
$T_d \sim 9$ K and $\beta_d = 1.67$.
The spectral index of each component is held constant
over the entire sky,
but the temperatures and relative amplitudes
are derived for each pixel following 
\cite{finkbeiner/etal:1999}.
The overall normalization of the combined dust emission
in Stokes $Q$ and $U$
is set to match the Planck polarized dust model at 353 GHz.
The fitted model assumes a superposition of
synchrotron, dust, and CMB emission
using a single modified blackbody 
(Eq. \ref{dust_greybody_eq}) for the dust,
for a total of 9 free parameters in each pixel
(Stokes $Q$ and $U$ amplitudes
for the CMB, synchrotron, and dust
plus the
synchrotron spectral index,
dust spectral index,
and dust temperature).
Figure \ref{2_greybody_cl}
shows the B-mode power spectrum
for the fitted CMB component.
Fitting two-component dust to a single dust component
creates CMB residuals 
with amplitude comparable to the inflationary B-modes
$r = 0.003$ at the recombination bump
and
$r = 0.006$ at the reionization bump.

\subsection{Multi-component Modified Blackbody}
Section $\S$4.3 demonstrates
that fitting two-temperature dust
using a single-temperature model
results in a bias $\Delta r > 0.003$.
Extending the fitted dust model to include a second component
reduces this bias.
As before, we quantify the bias
using an input sky 
with power-law synchrotron 
plus dust emission derived
from one of the toy models in Figure \ref{tau_vs_temp},
fitting the simulated sky
to a model with
CMB, synchrotron, and modified blackbody dust with two temperature components.
The model now has a total of 13 free parameters in each pixel:
8 parameters for the
Stokes $Q$ and $U$ amplitudes
of the
CMB, synchrotron, warm dust, and cold dust components,
plus an additional 5 free parameters 
(synchrotron spectral index,
warm dust temperature and spectral index,
cold dust temperature and spectral index)
to determine the foreground spectra,
which are assumed to be identical for Stokes $Q$ and $U$.
The resulting model
fits the two-component input exactly,
and reduces the bias in the CMB solution
for both the Gaussian and transient toy models.
The biases depend somewhat
on the detailed temperature distribution,
but are typically of order 
0.3 nK ($\Delta r \sim 1 \times 10^{-4}$)
for the two-Gaussian toy model
and
2 nK ($\Delta r \sim 8 \times 10^{-4}$)
for the transient heating model.

\subsection{Two-Level Systems and Running Spectral Index}
The models in $\S$4.2 -- 4.4
treat the far-IR dust spectrum as
a superposition of modified blackbody emission 
from components at different physical temperatures.
The far-IR emission of interstellar dust grains
is not well understood
and could plausibly utilize different physical mechanisms.
One example is the two-level system (TLS) emission model
developed by 
\citet{phillips:1972},
\citet{anderson/etal:1972},
and
\cite{bosch:1978}
and later extended to interstellar dust
\citep{
agladze/etal:1996,
meny/etal:2007,
paradis/etal:2011}.
Emission in this model is dominated by
acoustic oscillations in a disordered charge medium
and by
hopping relaxation
in a distribution of two-level tunneling states.
The temperature-independent acoustic oscillations
are the principal emission component at THz frequencies,
while hopping relaxation
adds an additional temperature-dependent component
to flatten the spectrum at millimeter wavelengths.

The TLS model has several attractive features.
It invokes solid-state physics 
to reproduce the observed anti-correlation
between the dust temperature and spectral index
derived from modified blackbody fits to the far-IR dust spectra
\citep{dupac/etal:2003,
desert/etal:2008,
bracco/etal:2011,
planck:2014:11}.
Emission from the tunneling component
reproduces the excess emission observed at millimeter wavelengths
without requiring a cold dust component.
Spectral fits 
using COBE/FIRAS data
show the TLS model to be a viable description of
far-IR emission from interstellar dust
\citep{paradis/etal:2011,odegard/etal:2016}.

A key feature of the TLS model
is the progressive flattening of the spectrum
toward lower frequencies.
It may be approximated as a
temperature-dependent ``running'' of the spectral index,
$\beta_d = \beta_d(T, \nu)$.
At dust temperatures $T_d = 20$ K 
typical for the diffuse interstellar cirrus,
the spectral index flattens from
$\beta_d = 2$ at THz frequencies
to $\beta_d \sim 1$ at 30 GHz
(see, e.g., Figure 6 of \citet{paradis/etal:2011}).

The spectral flattening at frequencies below 500 GHz
where CMB emission peaks
can bias multi-component fits 
relying on simpler models of the far-IR dust emission.
We quantify the bias in $r$ 
using a toy model of TLS dust.
The toy model approximates the TLS emission
using a running spectral index,
\begin{equation}
I_d(\nu) = \epsilon(\nu_d) B_\nu(T) 
\left( \frac{\nu}{\nu_d} \right)^{\beta_{d}(T, \nu)} .
\label{tls_approx}
\end{equation}
We adopt the input dust temperature 
for each sky pixel
using the Planck single-temperature modified blackbody model,
and interpolate $\beta_{d}(T, \nu)$
in frequency
using data from \citet{paradis/etal:2011}
appropriate for the dust temperature at that pixel.
The resulting spectrum is then normalized
to the Planck polarized dust model at 353 GHz.
We fit the superposed synchrotron and dust emission
to a model with CMB, synchrotron, and dust
using the same two-component modified blackbody
as $\S$4.4
and evaluate the resulting CMB component
to derive the bias in $r$.
The two-component modified blackbody model
reproduces the combined input sky well,
with fractional residuals
$\Delta I/I < 10^{-6}$
for all frequencies between 30 GHz and 500 GHz.
However, the best-fit sky includes a 
biased CMB component with typical bias
$\Delta r = 6 \times 10^{-3}$.

Note that the noiseless simulations
fit the foreground spectra independently for each pixel.
Actual observations include instrument noise
and typically fit some foreground parameters 
(temperature, amplitude)
within each pixel
while fitting others
(spectral index)
over larger regions of the sky.
The biases above thus represent an optimistic limiting case
where the signal to noise ratio allows
fitting 13 free parameters within each pixel.
Even for this optimistic case,
biases can be present at amplitudes
large compared to projected instrument sensitivities.

\section{Discussion}
Parametric modeling of the diffuse dust foreground
can bias estimates of CMB polarization
if the chosen dust model
cannot adequately represent the true emission spectrum.
For the commonly used one- or two-component modified blackbody models,
the bias can exceed levels 
$\Delta r \sim {\rm few} \times 10^{-3}$,
large compared to the sensitivities
anticipated for the next generation of CMB instrumentation
\citep{
kogut/etal:2011,		
matsumura/etal:2014,		
delabrouille:2015,		
abazajian/etal:2015}.		
Foreground bias at this level
would represent a statistically significant error
in derived cosmological parameters
and (depending on the cosmology)
could induce a false detection of the inflationary signal.

Several techniques may be employed to reduce this bias.
One method is to restrict the observing bands
to frequencies 
30 GHz $< \nu  <$ 250 GHz
where the CMB is brightest 
compared to the combined synchrotron and dust foregrounds,
thereby reducing the precision 
to which the foreground emission must be modeled.
Parametric fitting over a restricted  frequency range
also reduces the effect of un-modeled spectral curvature
present when the fitted spectral model
differs from the true foreground spectra.
However, the smaller foreground amplitudes
within the restricted frequency range
require correspondingly better sensitivity
in foreground-dominated channels
to identify and remove the foreground emission.

\begin{figure}[t]
\centerline{
\includegraphics[width=3.2in]{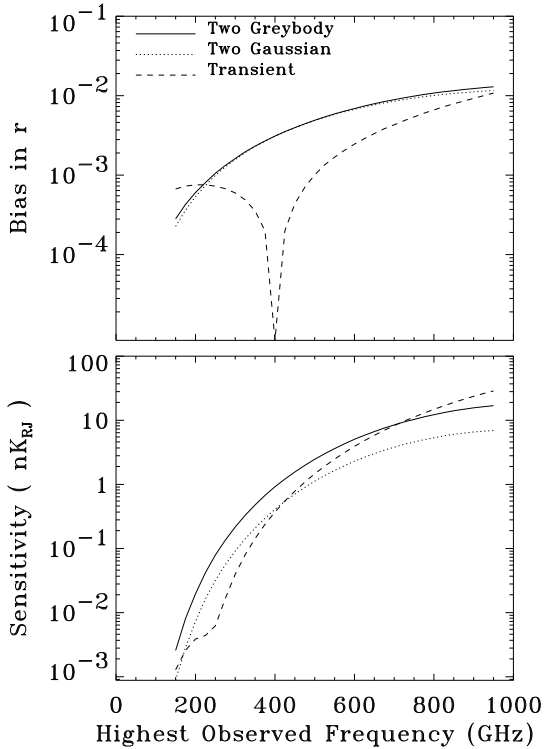} 
}
\caption{Effect of restricted observing frequencies
for parametric dust models.
(Top) Bias $\Delta r$ 
caused by fitting
a single-temperature modified blackbody dust model to a sky containing
dust emission from the
two-temperature, two-Gaussian, or transient heating
toy models (see text).
The best-fit value for $r$
changes sign when fitting the transient heating input.
(Bottom)
Residual between the input sky and best-fit model,
evaluated at the highest observed frequency.
}
\label{bias_vs_freq}
\end{figure}

We quantify this trade 
by fitting simulated skies
while varying the frequency range
over which the fit is performed.
The input sky consists of power-law synchrotron
with spectral index $-1.05$
plus dust specified by either the
two-temperature, two-Gaussian, or transient
toy models from $\S$4.4.
The fitted model includes CMB, synchrotron,
and single-temperature modified blackbody dust
for a total of 9 free parameters.
We fit the simulated sky
using 10 observing channels
spaced logarithmically in frequency
starting at 30 GHz,
and vary the highest observed frequency
from 150 GHz to 950 GHz
to show the effect of 
extending observations into the brighter dust foreground
at higher frequencies.

Figure \ref{bias_vs_freq} shows the results.
For the two-temperature and two-Gaussian input models,
the simulations show the expected pattern
with the bias $\Delta r$
monotonically decreasing
as the frequency range is restricted.
Fitting a single modified blackbody to a toy model
with a high-temperature tail
shows a different pattern.
When the fit is restricted to frequencies below 400 GHz,
the single-component model under-estimates the dust emission
and over-estimates the CMB.
When the frequency range is extended to the higher frequencies
where the warmer components dominate,
the single-component model over-estimates the dust emission
and under-estimates the CMB.
The residual CMB term thus changes sign,
creating a null for the special case
where the frequency coverage extends just high enough
that the dust residuals cancel.
Note that,
for the transient heating toy model,
restricting the frequency coverage
below 400 GHz actually
{\it increases} the bias in $r$.

A second consideration for restricted frequency coverage
is sensitivity.
As the frequency coverage is restricted,
the parametric models better reconstruct
the superposed emission from the CMB and combined foregrounds.
This is not the same as accurately reconstructing 
the CMB signal itself.
For the noiseless simulations in this paper,
we define a goodness-of-fit statistic $\Gamma$
using the rms difference between the
intensities of the input sky and the best-fit model,
summed over observing channels
(Eq. \ref{resid_eq}).
This fractional residual
varies from
$\Gamma = 10^{-5}$
for fits restricted to frequencies $\nu < 200$ GHz
to
$\Gamma = 10^{-3}$
for fits extending to $\nu < 950$ GHz
and is nearly identical for all three toy models considered.

The toy models evaluated above
yield non-trivial bias in $r$
despite fitting the combined sky emission to sub-percent precision.
Since the functional form of far-IR dust emission 
is not known {\it a priori},
distinguishing between biased and unbiased fits
requires channel sensitivities
comparable to the residuals
derived from plausible models of dust emission.
The design of CMB missions
typically includes one or more
``guard channels''
at higher frequencies where dust emission dominates.
The bottom panel of Figure \ref{bias_vs_freq}
shows the guard channel sensitivity 
required to distinguish
these toy models,
\begin{equation}
\delta S = \Gamma ~I_{\rm input} \vert_{\nu = \nu_{\rm max}} ,
\label{sensitivity_def}
\end{equation}
evaluated as a function of the highest observing frequency.
Since the highest observing frequency may extend
to frequencies beyond the Wien cutoff in the CMB intensity,
we present the sensitivity $\delta S$
in units of Rayleigh-Jeans temperature,
\begin{equation}
\delta T_{\rm RJ} = \delta S ~\frac{\lambda^2}{2k}
\label{rj_def}
\end{equation}
where
$k$ is Boltzmann's constant and
$\lambda$ is the observing wavelength.

Figure \ref{bias_vs_freq} illustrates the challenges
for observations with restricted frequency coverage.
A nine-parameter fit 
limited to observations at frequencies $\nu < 250$ GHz
most accessible to ground-based observations
yields biases $\Delta r~>~10^{-3}$
despite fitting the combined sky emission 
to precision $\Gamma \sim 10^{-5}$.
A statistical detection of this bias
would require sub-nK channel sensitivities.

At least two solutions are possible.
Fits with additional free parameters
can better model the (unknown) dust spectrum,
reducing the bias in the fitted CMB component.
A 13-parameter model reduces the bias substantially
($\S$4.4)
but would require at least 14 observing channels.
Fitting multiple temperature components
within a restricted frequency range
may additionally lead to parameter degeneracies.

An alternative is to include observations at higher frequencies.
Extending the observations to THz frequencies 
where the dust emission peaks
provides discrimination among various dust models.
For example, extending observations to 3 THz
increases the residual between the input sky and best-fit model to 
400 Jy sr$^{-1}$ for the TLS model
and
over 2000 Jy sr$^{-1}$ for the transient heating model.
Differences at this level
are readily detectable,
providing statistical evidence
for a poorly fitted model.

\begin{figure}[t]
\centerline{ 
\includegraphics[width=2.7in,angle=90]{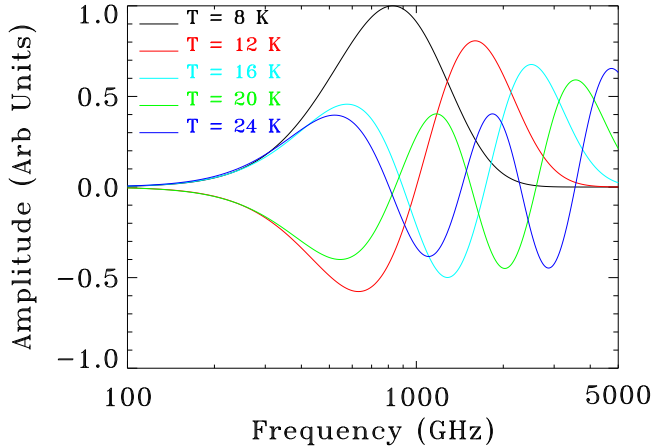}
}
\caption{
Dust spectral components
corresponding to different physical temperatures,
derived from Gram-Schmidt orthogonalization
of the individual Planck spectra at each temperature.
The minimal spectral content at frequencies below 800 GHz
makes it difficult for observations at low frequencies
to distinguish multiple temperature components:
most of the spectral content lies at frequencies above 1 THz.
}
\label{orthogonal_spectra}
\end{figure}

Observations at frequencies 
at or above the peak of the dust spectrum
can additionally provide information 
on the actual dust temperature distribution,
further discriminating between candidate dust models.
Emission from dust components at successively higher temperatures
will peak at successively higher frequencies 
(the well-known Wien displacement law);
however,
the superposed emission from multiple temperature components
will typically show a single broad peak.
One indicator of the temperature information
within the superposed spectrum
may be shown by Gram-Schmidt orthogonalization.
We begin with a Planck spectrum
at some minimum temperature $T_{\rm min}$,
and
construct orthogonal linear combinations
from Planck spectra 
at successively higher temperatures
to some maximum $T_{\rm max}$.
Figure \ref{orthogonal_spectra}
shows the resulting orthogonal components
for temperatures
8~K $< ~T ~ <$ 24 K,
evaluated at frequencies $\nu < 5$ THz.
There is relatively little spectral content
at frequencies below 800 GHz,
but well-separated peaks in the component spectra
at higher frequencies.

Another way to evaluate the ability of THz observations
to constrain the dust temperature distribution
is to note that
emission on the Rayleigh-Jeans side of the emission peak
shows flatter spectral dependence
compared to the exponential Wien cutoff at higher frequencies.
Dust at temperatures $T_d > 8$ K
will peak at frequencies above 800 GHz,
so it is not surprising to find most of the spectral content
at frequencies above 800 GHz.

\section{Summary}
Far-IR dust emission is complex.
Variation in the chemical or physical composition of the dust grains
or in the local stellar radiation field incident on each grain
will drive variation in the physical temperature of the grains.
Transient heating of individual grains
drives further temperature variation,
even for a homogeneous dust population
illuminated by a uniform radiation field.
Phenomenological models treating the far-IR emission spectrum
as modified blackbody emission
at one or two well-defined temperatures
can only approximate the dust spectrum.
Systematic differences between the actual emission spectrum
and the simpler parametric forms 
used to model the emission
will then bias both the recovered CMB and dust parameters.

We use toy models of far-IR dust emission
to estimate the accuracy to which
the polarization of the cosmic microwave background
can be recovered 
if the parametric form chosen for the dust component
of a multi-frequency fit
differs from the actual dust emission.
We generate noiseless simulations
combining power-law synchrotron emission
with toy models of dust emission,
evaluate the combined emission at a set of observing frequencies
spanning the minimum of the foreground spectra,
then fit the resulting multi-frequency spectra
to a superposition of synchrotron, dust, and CMB emission
to derive the bias in CMB polarization.
Differences between the input sky simulation
and the parametric models used to fit the dust component
yield CMB residuals comparable to or larger than
the sensitivities expected for the next generation of CMB missions,
despite fitting the combined sky emission
to a precision $\Delta I / I$ of $10^{-3}$ or better.

The amplitude of the CMB residual
depends on the complexity of the input dust model,
the number of parameters used by the output dust model,
and the frequency range over which the fit is performed.
Although dust at millimeter wavelengths is often modeled 
as a modified blackbody at a single, well-defined temperature,
the spectrum is known to be more complex.
Observations at higher frequencies
show excess emission 
with respect to the single-temperature modified blackbody dust model.
The excess is consistent
with emission from a second, colder dust component.
Simulations with input skies containing
either the two-component modified blackbody dust model
or
emission from extended temperature distributions
but fit to a nine-parameter model with 
a single-temperature modified blackbody dust component
consistently yield CMB residuals $\Delta r > 0.003$
for the inflationary B-mode amplitude.
Restricting observations to frequencies
30 GHz $< ~\nu ~ <$ 250 GHz
where the CMB is brightest relative to the foregrounds
reduces the bias for some of the two-component input model,
but actually increases the bias 
when the input model 
assumes a continuous distribution of dust emissivity 
extending to higher temperatures.

Adding additional free parameters to the output dust model
reduces the CMB bias in certain cases.
A 13-parameter output model fitting two modified blackbody dust components
reproduces the two-component input model exactly,
and reduces CMB bias by an order of magnitude
for input models with extended temperature distributions.
However, the 13-parameter model
fails when the input sky contains dust with a running spectral index.
Such a running index is consistent with observations
and is expected for dust emission originating from 
two-level systems within the disordered medium of interstellar dust grains.
A 13-parameter fit to the TLS model
showed CMB residuals
$\Delta r = 6 \times 10^{-3}$
despite fitting the combined CMB$+$foreground emission
at frequencies $\nu < 500$ GHz
to precision $\Delta I / I < 10^{-6}$.

Observations at THz frequencies 
surrounding the peak of dust emission
can mitigate these biases.
Extending parametric fits over a broader frequency range
increases the residuals
when the parametric form of the fitted model
poorly matches the actual sky.
At THz frequencies,
the output-input dust residual for the toy models
had readily detectable values 400--2000 Jy sr$^{-1}$,
allowing statistical detection 
of poorly fitted models.
Observations at THz frequencies
would also allow direct determination
of the dust temperature profile,
informing fits at lower frequencies.


\end{document}